\documentclass[
preprint,
prd,
aps,
eqsecnum,
amsmath,
amssymb,
]{revtex4-1}
\def\Tr{\mbox{Tr}\,}

\usepackage{graphicx}
\usepackage{color}
\usepackage{float}
\usepackage{subfloat}
\usepackage{subfigure}

\usepackage{hyperref}

\begin{document}
\title{Masses of physical scalars in two Higgs doublet models} 
\author{Ambalika Biswas}\email{ambalika12t@boson.bose.res.in}
\author{Amitabha Lahiri}\email{amitabha@bose.res.in}
\affiliation{S. N. Bose National Centre For Basic Sciences,\\
 Block JD, Sector III, Salt Lake, Kolkata 700098, INDIA}
\date{\today}

\begin{abstract}
  We find bounds on scalar masses resulting from a criterion of
  naturalness, in a broad class of two Higgs doublet models
  (2HDMs). Specifically, we assume the cancellation of quadratic
  divergences in what are called the type I, type II,
  lepton-specific and flipped 2HDMs, with an additional U(1)
  symmetry. This results in a set of relations among masses of the
  physical scalars and coupling constants, a generalization of the
  Veltman conditions of the Standard Model. Assuming that the
  lighter $CP$-even neutral Higgs particle is the observed scalar
  particle of mass $\sim$125 GeV, and imposing further the
  constraints from the electroweak T-parameter, stability, and
  perturbative unitarity, we calculate the range of the mass of
  each of the remaining physical scalars.

\end{abstract}

\maketitle

\section{Introduction}
With the discovery of a 125~GeV neutral scalar
boson~\cite{Aad:2012tfa, Chatrchyan:2012ufa}, the menagerie of
fundamental particles in the Standard Model appears to be
complete. Some questions still remain unanswered, including the
origins of neutrino mass and dark matter, keeping the door open for
physics beyond the Standard Model. Among the simplest extensions of
the Standard Model are two Higgs doublet models (2HDMs) (for a
recent review see~\cite{Branco:2011iw}).  Originally motivated by
supersymmetry, where a second Higgs doublet is essential, 2HDMs
have also been studied in several other contexts. Peccei-Quinn
symmetry~\cite{Peccei:1977hh, Peccei:1977ur} solves the strong CP
problem, but must be spontaneously broken. The corresponding
Goldstone boson is the axion, which can be a combination of the
phases of two Higgs doublets. Models of baryogenesis often involve
2HDMs~\cite{Turok:1990zg} because their mass spectrum can be
adjusted to produce CP violation, both explicit and spontaneous.
Another motivation, one that is important to us, is their use in
models of dark matter~\cite{Ma:2006km, Ma:2008uza,
  Barbieri:2006dq}.  These models are the inert doublet models, so
called because one of the Higgs doublets does not couple to the
fermions. Of the 2HDMs we will consider, the Yukawa couplings of 
one model (type I) approach the inert doublet model for large 
values of the ratio of the vacuum expectation values (VEVs) of 
the two Higgs fields. The other models also have small couplings 
to one or more types of fermions in that limit.

In this paper we consider 2HDMs with a softly broken global
U(1) symmetry~\cite{Peccei:1977hh, Ferreira:2009jb}, with the 
parameters chosen so as to make the 2HDM `SM-like'. We choose 
the fermion transformations under this U(1) symmetry, and impose 
a naturalness condition of vanishing quadratic divergences on the 
scalar sector of the models. Using additional restrictions coming
from partial wave unitarity, vacuum stability, and the $T$
parameter measuring `new physics', and assuming that the 
lighter CP-even Higgs particle in the 2HDMs is the one observed at the 
Large Hadron Collider (LHC), we find bounds on the masses of
the additional scalar particles for each of the 2HDMs.

We will work with the  scalar potential~\cite{Lee:1973iz,
  Gunion:1989we} 
\begin{align}
V &=
\lambda_{1}\left(|\Phi_{1}|^2 
- \frac{v_{1}^{2}}{2}\right)^{2} + 
\lambda_{2}\left(|\Phi_{2}|^2 
- \frac{v_{2}^{2}}{2}\right)^{2}     
 \nonumber \\
 & \quad 
+\lambda_3\left(|\Phi_1|^2 + |\Phi_2|^2
-\frac{v_{1}^{2} +  v_{2}^{2}}{2}\right)^{2}  
\nonumber \\ 
& \quad
+\lambda_{4}\left(|\Phi_{1}|^2 |\Phi_{2}|^2 
- |\Phi_{1}^{\dagger}\Phi_{2}|^2\right) 
 \nonumber \\
 & \quad 
 + \lambda_5\left|\Phi_1^\dagger\Phi_2 -  \frac{v_1v_2}{2}\right|^2\,,
\label{2HDM.potential}
\end{align}
with real $\lambda_i$.  This potential is
invariant under the symmetry 
 $\Phi_1 \to e^{i\theta}\Phi_1\,, \Phi_2 \to \Phi_2\,,$
except for a soft breaking term $\lambda_5 v_1v_2
\Re(\Phi_{1}^{\dagger} \Phi_{2})\,.$ Additional dimension-4 terms,
including one allowed by a softly broken $Z_2$ symmetry~\cite{Gunion:1992hs}
are also set to zero by this U(1) symmetry. 

The  scalar doublets are parametrized as
\begin{equation}
\Phi_{i}=
\left(
\begin{array}{c}
w_{i}^{+}(x)\\
\dfrac{v_{i}+h_{i}(x)+iz_{i}(x)}{\sqrt{2}}
\end{array}
\right)\,, \qquad i=1,2
\label{doublet.vev}
\end{equation}
where the VEVs $v_i$ may be taken to be real and positive without
any loss of generality.  Three of these fields get ``eaten'' by the
$W^{\pm}$ and $Z^{0}$ gauge bosons; the remaining five are physical
scalar (Higgs) fields. There is a pair of charged scalars denoted
by $\xi^{\pm}$, two neutral CP even scalars $H$ and $h$\,,
and one CP odd pseudoscalar denoted by
$A$. With
\begin{equation}
\tan\beta=\frac{v_{2}}{v_{1}}\,, \label{tanbeta.def}
\end{equation}
these fields are given by the combinations
\begin{equation}
\left(
\begin{array}{c}
\omega^{\pm}\\
\xi^{\pm}
\end{array}
\right)=\left(
\begin{array}{rcl}
c_{\beta}& s_{\beta}\\
-s_{\beta}& c_{\beta}
\end{array}
\right) \left(
\begin{array}{c}
w_{1}^{\pm}\\
w_{2}^{\pm}
\end{array}
\right),
\label{redef.charged}
\end{equation}
\begin{equation}
\left(
\begin{array}{c}
\zeta\\
A
\end{array}
\right)=\left(
\begin{array}{rcl}
c_{\beta}& s_{\beta}\\
-s_{\beta}& c_{\beta}
\end{array}
\right) \left(
\begin{array}{c}
z_{1}\\
z_{2}
\end{array}
\right),
\label{redef.axial}
\end{equation}
\begin{equation}
\left(
\begin{array}{c}
H\\
h
\end{array}
\right)=\left(
\begin{array}{rcl}
c_{\alpha}& s_{\alpha}\\
-s_{\alpha}& c_{\alpha}
\end{array}
\right) \left(
\begin{array}{c}
h_{1}\\
h_{2}
\end{array}
\right),
\label{redef.higgs}
\end{equation}
where  $c_{\alpha}\equiv \cos\alpha$ etc.

If we rotated $h_{1} - h_{2}$ fields by the angle $\beta$, 
\begin{equation}
\left(
\begin{array}{c}
H^{0}\\
R
\end{array}
\right) = \left(
\begin{array}{rcl}
c_{\beta}& s_{\beta}\\
-s_{\beta}& c_{\beta}
\end{array}
\right) \left(
\begin{array}{c}
h_{1}\\
h_{2}
\end{array}
\right),
\label{Hbeta}
\end{equation}
we would find that $H^{0}$ has exactly the Standard Model Higgs
couplings with the fermions and gauge bosons~\cite{Branco:1996bq, 
	Gunion:2002zf}.
The physical scalar $h$ is related to $H^{0}$ and $R$ via 
\begin{equation}
h=\sin(\beta-\alpha)H^{0}+ \cos(\beta-\alpha)R\,.
\label{decoupling.1}
\end{equation}
Thus in order for $h$ to be the Higgs boson of the Standard Model,
we require $\sin(\beta-\alpha)\approx 1\,,$ which has been called the 
SM-like or alignment limit~\cite{Ferreira:2014naa}. Accordingly, we will 
assume $\beta-\alpha = \tfrac{\pi}{2}\,$ in the rest of this paper.

\section{Veltman Conditions}
The scalar masses get quadratically divergent contributions which 
require a fine-tuning of parameters. We thus impose naturalness
conditions, a generalization of the Veltman conditions for the 
Standard Model, that these contributions cancel~\cite{Veltman:1980mj}.
The resulting masses and couplings should not then require 
fine-tuning.

The Yukawa potential for the 2HDMs is of the form
\begin{equation}
{\cal L}_{Y}= \sum_{i=1,2}
\left[- \bar{l}_L\Phi_iG_{e}^i e_R 
- \bar{Q}_L \tilde{\Phi}_{i}
G_{u}^{i}u_{R}
- \bar{Q}_{L}\Phi_{i}G_{d}^{i} d_{R} + h.c.\right]
\,,
\label{Lag.Yukawa2}
\end{equation}
where $l_L\,, Q_L$ are 3-vectors of isodoublets in the space of
generations, $e_R\,, u_R\,, d_R$ are 3-vectors of singlets, $G^1_e$
etc. are complex $3\times 3$ matrices in generation space
containing the Yukawa coupling constants, and
$\tilde\Phi_i=i\tau_2\Phi_i^*\,.$

Cancellation of quadratic divergences in the scalar masses gives
rise to four mass relations, which we may call the Veltman
conditions for the 2HDMs being considered~\cite{Newton:1993xc},
%
\begin{align}
2\Tr G_{e}^{1}G_{e}^{1\dagger} + 6\Tr G_{u}^{1\dagger}G_{u}^{1} 
+ 6\Tr G_{d}^{1}G_{d}^{1\dagger} &= 
\frac{9}{4}g^{2}+\frac{3}{4}g^{\prime 2}+6\lambda_{1}
+10\lambda_{3}+\lambda_{4} + \lambda_5 \,, 
\label{VC.vc1}\\
2\Tr G_{e}^{2}G_{e}^{2\dagger} + 6\Tr G_{u}^{2\dagger}G_{u}^{2}
+ 6\Tr G_{d}^{2}G_{d}^{2\dagger} &= 
\frac{9}{4}g^{2}+\frac{3}{4}g^{\prime 2}+6\lambda_{2}
+10\lambda_{3}+\lambda_{4}+ \lambda_5\,,
\label{VC.vc2} \\
2\Tr G_{e}^{1}G_{e}^{2\dagger} + 6\Tr G_{u}^{1\dagger}G_{u}^{2}
+ 6\Tr G_{d}^{1}G_{d}^{2\dagger} &=0\,,
\label{VC.vc3}
\end{align}
%
where $g, g'$ are the $SU(2)$ and $U(1)_Y$ coupling constants.  A
fourth equation is the complex conjugate of the third one.  
As we will see below, the last equation vanish identically for all
the 2HDMs we consider. The mass relations come from the first two
equations above.

When the fermions are in mass eigenstates, the Yukawa
matrices are automatically diagonal if there is only one Higgs
doublet as in the Standard Model, so there is no FCNC at the tree
level. But in the presence of a second scalar doublet, the two
Yukawa matrices will not be simultaneously diagonalizable in
general, and thus the Yukawa couplings will not be flavor
diagonal. Neutral Higgs scalars will mediate FCNCs.  The necessary and
sufficient condition for the absence of FCNC at tree level is that
all fermions of a given charge and helicity transform according to
the same irreducible representation of SU(2), corresponding to the
same eigenvalue of $T_{3}\,,$ and that a basis exists in which they
receive their contributions in the mass matrix from a single
source~\cite{Glashow:1976nt, Paschos:1976ay}.

For the fermions of the Standard Model,
this theorem implies that all right-handed singlets of a given
charge must couple to the same Higgs doublet. We will ensure this
using the global U(1) symmetry mentioned earlier, which generalizes a
$Z_2$ symmetry more commonly employed for this purpose.
The left handed fermion doublets remain unchanged under this
symmetry, $Q_L \to Q_L\,, l_L \to l_L\,.$ The transformations of
right handed fermion singlets determine the type of 2HDM. There are
four such possibilities, which may be identified by the
right-handed fields which transform under the U(1): type I (none),
type II ($d_{R}\rightarrow e^{-i\theta}d_{R}\,, e_{R}\rightarrow
e^{-i\theta}e_{R}$)\,, lepton specific ($e_{R}\rightarrow
e^{-i\theta}e_{R}$)\,, flipped ($d_{R}\rightarrow
e^{-i\theta}d_{R}$)\,. 
We note in passing that another way of avoiding FCNCs at tree level
is by aligning the Yukawa and mass matrices in flavor
space~\cite{Pich:2009sp}. However, only these four 2HDMs admit 
symmetries such as the U(1)~\cite{Ferreira:2010xe}.

The fermion mass matrix is diagonalized by independent unitary
transformations on the left and right-handed fermion fields. In any
of the 2HDMs, either $G_{1f}$ or $G_{2f}$ vanish for each fermion
type $f\,.$ For example, in the Type II model $\Phi_{1}$ couples to
down-type quarks and charged leptons,  while $\Phi_{2}$ couples to
up-type quarks, so $G_{2e}= G_{2d}= G_{1u}=0\,.$ Thus
Eq.~(\ref{VC.vc3}) is automatically satisfied in each 2HDM.
The non-vanishing Yukawa matrices are related to the fermion masses
by~\cite{Newton:1993xc}
\begin{align}
\Tr[G_{1f}^{\dagger}G_{1f}] &=\frac{2}{v^2 \cos^{2}\beta} \sum
m_f^2\,, \label{Yukawa.1}\\  
\Tr[G_{2f}^{\dagger}G_{2f}] &=\frac{2}{v^2\sin^{2}\beta} \sum
m_f^2\,, \label{Yukawa.2} 
\end{align}
where $f$ stands for charged leptons, up-type
quarks, or down-type quarks, and the sum is taken over generations.

In order to rewrite the Veltman conditions in terms of the known
masses, we first note that in the alignment limit and with the
global U(1) symmetry, the independent parameters in the scalar
potential may be taken to be the masses $m_h\,, m_H\,, m_\xi\,,$
the angle $\beta\,,$ the electroweak VEV $v=\sqrt{v_{1}^{2} 
	+v_{2}^{2}} \,,$ and the constant
$\lambda_5\,.$ The $\lambda_i$ are related to these parameters
by~\cite{Akeroyd:2000wc}
\begin{align}
\lambda_{1} &= \frac{1}{2v^{2} c^2_\beta}
m_{H}^{2}
-\frac{\lambda_{5}}{4}(\tan^{2}\beta-1)\,, \label{mass.lambda1}\\  
\lambda_{2} &= \frac{1}{2v^{2}s^{2}_{\beta}}
m_{H}^{2}
-\frac{\lambda_{5}}{4}\left(\frac{1}{\tan^{2}\beta} -1\right)\,,
\label{mass.lambda2}\\ 
\lambda_{3} &= -\frac{1}{2v^{2}}(m_{H}^{2}-m_{h}^{2})
-\frac{\lambda_{5}}{4}\,,\label{mass.lambda3}\\  
\lambda_{4} &= \frac{2}{v^{2}}m_{\xi}^{2}\,,
\qquad \lambda_5 = \frac{2}{v^{2}}m_{A}^{2}\,.
\label{mass.lambda5}
\end{align}
Inserting Eq.~(\ref{Yukawa.1}) --- Eq.~(\ref{mass.lambda5}) into
Eq.~(\ref{VC.vc1}) and Eq.~(\ref{VC.vc2}), we get the Veltman
conditions in terms of the physical particle masses. These are shown 
in Table~\ref{table.VC}. The Yukawa matrices which vanish in each 
model are listed in the second column. We note here that 
although naturalness conditions in specific
2HDMs have been studied earlier on a few 
occasions~\cite{Grzadkowski:2009iz, Jora:2013opa}, 
they were not done in the SM-like scenario, 
nor expressed in terms of the physical masses
for the different types as in here.

%
\begin{table}[tbhp]
\begin{tabular}{||c|c|c|c||}
\hline
Model\; &\; zero Yukawa\; 
&  VC1 & VC2 \\
\hline
&&
 $6M_W^{2}+3M_Z^{2}
+ 5 m_{h}^{2} + 2m_{\xi}^{2}  $ 
&
$6M_W^{2}+3M_Z^{2} + 5 m_{h}^{2}
+ 2m_{\xi}^{2}  $\;\\ && 
\; $+\, m_{H}^{2}(3 \tan^2\beta - 2)
-\frac{3v^2}{2}
\lambda_{5} \tan^{2}\beta =$ \;
&
\; $+\, m_{H}^{2}\left(3\cot^2\beta - 2\right)
- \frac{3 v^2}{2}\lambda_{5}\cot^{2}\beta=$
\; \\
\hline
Type I &  $G_{1e}\,,  G_{1d}\,, G_{1u}\,$ & 0 &  
$4\left[\sum 
m_e^2 + 3 \sum m_u^2 + 3
  \sum m_d^2\right]\csc^{2}\beta\, $\\
Type II &  $G_{2e}\,, G_{2d}\,, G_{1u}\,$ &
 $ 4 \left[ \sum m_e^2 + 3\sum m_d^2 \right]\sec^2\beta\,$ 
&  
$12 \sum m^2_u \csc^2\beta\,$ \\
LS &  $G_{2e}\,, G_{1d}\,, G_{1u}\,$ &
$4\sum m_e^2 \sec^{2}\beta\,$
&
$12\left[ \sum m_u^2 + \sum m_d^2 \right] \csc^{2}\beta\,$ \\
Flipped & $G_{1e}\,,G_{2d} \,, G_{1u}\,$ &
$12 \sum m_d^2 \sec^2\beta\,$ 
& 
$ 4\left[\sum m_e^2 + 3 \sum m_{u}^{2} \right] \csc^{2}\beta\,$ \\
\hline
\end{tabular}
\caption{Veltman conditions for the different 2HDMs}
\label{table.VC}
\end{table}
%

\section{Bounds on the masses of heavy and charged scalars}\label{bounds}
We now display our main results, the bounds we have 
obtained for the masses of the heavy and charged Higgs 
particles. We will assume that the $h$ particle 
is the one that has been observed at the LHC, so that 
$m_{h}=125$ GeV, and $v= 246$ GeV.
 Let us consider the example of the type II model
to explain our derivation of the bounds.

Since we want the bounds on $m_H$ and
$m_\xi\,,$ let us rewrite VC1 and VC2 for the type II model 
in a convenient form, 
\begin{align}
m_{H}^{2}\left(3\tan^{2}\beta - 2\right) 
+ 2 m_{\xi}^{2}
= & 4\left[ \sum m_e^2 + 3 \sum m_d^2\right]\sec^{2}\beta
- 6M_W^{2} - 3M_Z^{2}
- 5 m_{h}^{2} + \lambda_{5}\frac{3v^{2}}{2}\tan^{2}\beta
\,,
\label{typeII.VC1} \\
m_{H}^{2}\left(3\cot^{2}\beta - 2\right) 
+ 2 m_{\xi}^{2}
= &12 \sum m_u^2 \csc^{2}\beta - 6M_W^{2} - 3M_Z^{2}
- 5 m_{h}^{2} + \lambda_{5}\frac{3v^{2}}{2}\cot^{2}\beta
\,.   
\label{typeII.VC2}
\end{align}
On the right hand side of either equation, all but the last term are
experimentally known. The $U(1)$ symmetry 
implies that $\lambda_5 > 0\,,$ and we impose the
restriction of~$\vert \lambda_{i} \vert \leq 4\pi$\,
based on the validity of perturbativity. Comparing with 
Eq.~(\ref{mass.lambda5}), we see that this last puts a 
restriction $m_A \lesssim 617$GeV. 

For a fixed value of $\tan\beta\,,$ we plot both equations on the 
$m_H-m_\xi$ plane for various values of $\lambda_5\,.$ The point
where the two curves cross for a given value of $\lambda_5$,
is an allowed value of the pair $(m_H\,, m_\xi)\,.$

We can restrict the allowed range of the masses even further by
imposing constraints coming from stability, perturbative unitarity,
and the oblique electroweak $T$-parameter. Conditions for
stability, i.e. for the scalar potential being bounded from below,
were examined in~\cite{Sher:1988mj, Gunion:2002zf, Branco:2011iw},
and found to provide lower bounds on certain combinations of the
quartic couplings $\lambda_i\,.$ On the other hand, the requirement
of perturbative unitarity translates into upper limits on
combinations of the $\lambda_i$\,, which for two-Higgs models
have been derived by many authors~\cite{Maalampi:1991fb,
  Kanemura:1993hm, Akeroyd:2000wc, Horejsi:2005da}.
One condition coming from perturbative unitarity is 
\begin{equation}
\left| 3(\lambda_{1}+\lambda_{2}+2\lambda_{3})\pm \sqrt{9(\lambda_{1}
	-\lambda_{2})^{2}+(4\lambda_{3}+\lambda_{4}+\lambda_{5})^{2}}\right|
 \ \leq 16\pi
\label{unitarity.1}
\end{equation}
Stability provides the inequalities 
\begin{equation}
\lambda_{1}+\lambda_{3}>0\,, \qquad
\lambda_{2}+\lambda_{3}>0\,,
\end{equation}
so that we can write Eq.~(\ref{unitarity.1}) as $|A \pm B| \leq 16\pi\,,$ with 
$A\,, B \geq 0\,.$ It then follows that
%
\begin{equation}
0 \leq \lambda_{1}+\lambda_{2}+2\lambda_{3} \leq \frac{16\pi}{3}\,.
\end{equation}
In terms of the scalar masses, this reads
\begin{equation}
0 < (m_{H}^{2}-m_{A}^{2})(\tan^{2}\beta + \cot^{2}\beta) 
+2m_{h}^{2} < \frac{32\pi v^{2}}{3}.
\label{mass.ineq1}
\end{equation}

\begin{figure}[htbp]
\includegraphics[height=0.45\columnwidth=width]{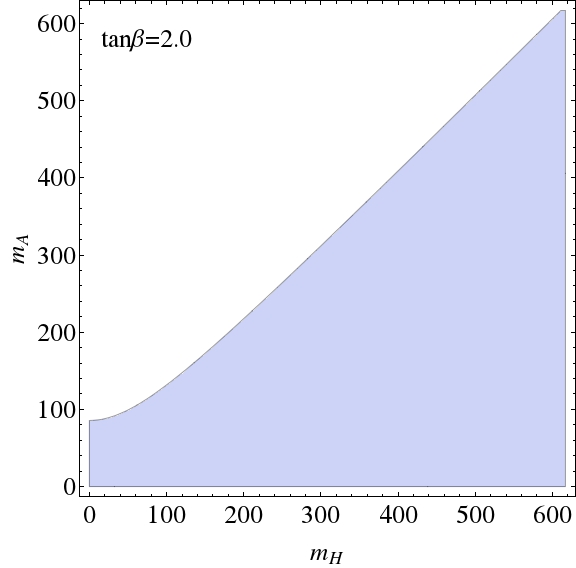} \;
\includegraphics[height=0.45\columnwidth=width]{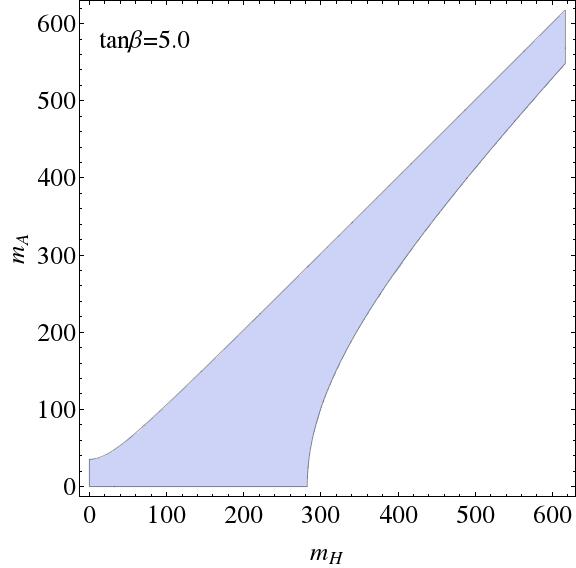} \\
\medskip
\includegraphics[height=0.45\columnwidth=width]{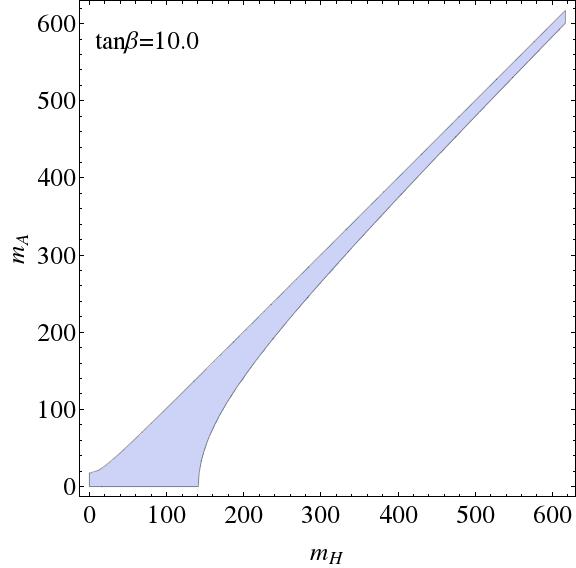} \;
\includegraphics[height=0.45\columnwidth=width]{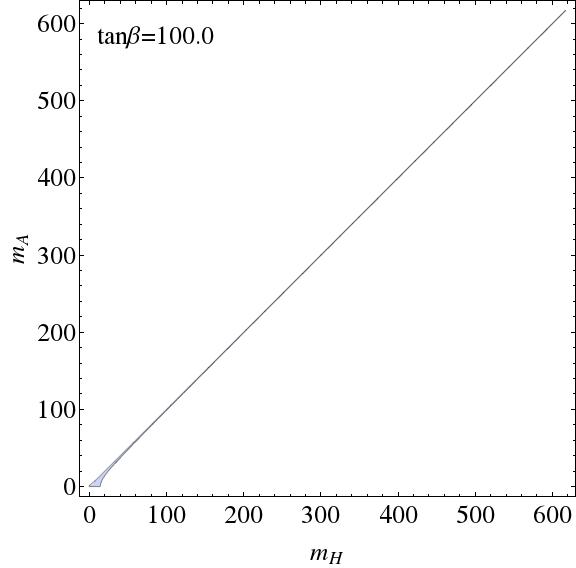}
\caption{Degeneracy of $m_{H} - m_{A}$ (in GeV) for progressively 
increasing $\tan\beta$\,. The condition $|\lambda_i|\leq 4\pi$ restricts
$m_A \lesssim$ 617 GeV.}
\label{fig.mAmH}
\end{figure}

For $\tan\beta \gg 1$, this inequality implies that
$m_{H}$ and $m_{A}$ are almost degenerate, a result also 
found in~\cite{Bhattacharyya:2013rya}. In Fig.~\ref{fig.mAmH}
we have shown this degeneracy by plotting $m_A$ against $m_H$
for different values of $\tan\beta\,.$ It is easy to see from the plots 
that the degeneracy is more pronounced at higher values of $m_A$
for any value of $\tan\beta\,.$  For these plots we have used the 
perturbativity condition $|\lambda_i| \leq 4\pi\,,$ which restricts 
$m_A \lesssim$ 617 GeV. 

We will also need another inequality which follows from the condition 
\begin{equation}
\left| 2\lambda_{3}+\lambda_{4} \right|\, \leq 16\pi
\end{equation}
required for perturbative unitarity. Substituting the mass relations
Eq.~(\ref{mass.lambda3}) and~(\ref{mass.lambda5}) into this, we get
\begin{equation}
\left\vert 2m_{\xi}^{2}-m_{H}^{2}-m_{A}^{2}+m_{h}^{2}
\right\vert \leq 16\pi v^{2}\,.
\label{mass.ineq2}
\end{equation}
Next we take into account the oblique
parameter $T$ for the 2HDMs, which has the
expression~\cite{He:2001tp, Grimus:2007if}
\begin{equation}
T=\frac{1}{16\pi \sin^{2}\theta_{W}M_{W}^{2}}
\left[F(m_{\xi}^{2},m_{H}^{2})+F(m_{\xi}^{2},
m_{A}^{2})-F(m_{H}^{2},m_{A}^{2})\right],\label{Tdef}  
\end{equation}
with
\begin{equation}
F(x,y) =
\left\{
\begin{array}{lr}
\frac{x+y}{2}-\frac{xy}{x-y}\ln\frac{x}{y}\,, \qquad & x\neq y\\ 
0&x=y
\end{array}
\right.
\label{Fdef}
\end{equation}
The $T$ parameter is constrained by the global fit to precision
electroweak data to be~\cite{Baak:2013ppa}
\begin{equation}
T=0.05\pm 0.12.
\label{Tvalue}
\end{equation}

\begin{figure}
\subfigure[]{
 \includegraphics[height=0.3\columnwidth, width = 0.45\columnwidth]{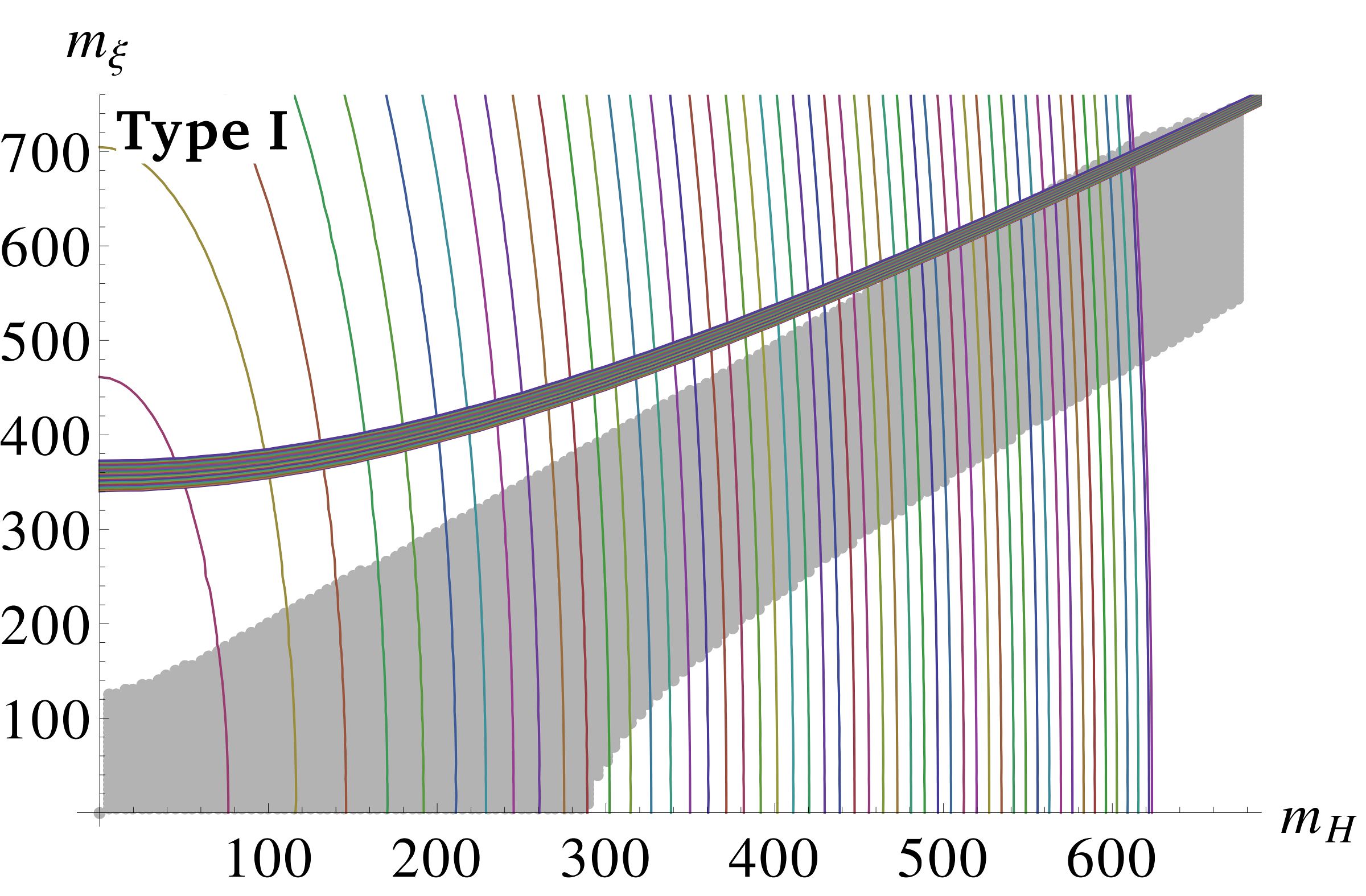} 
 \label{typeI}}
\subfigure[]{
 \includegraphics[height=0.3\columnwidth, width = 0.45\columnwidth]{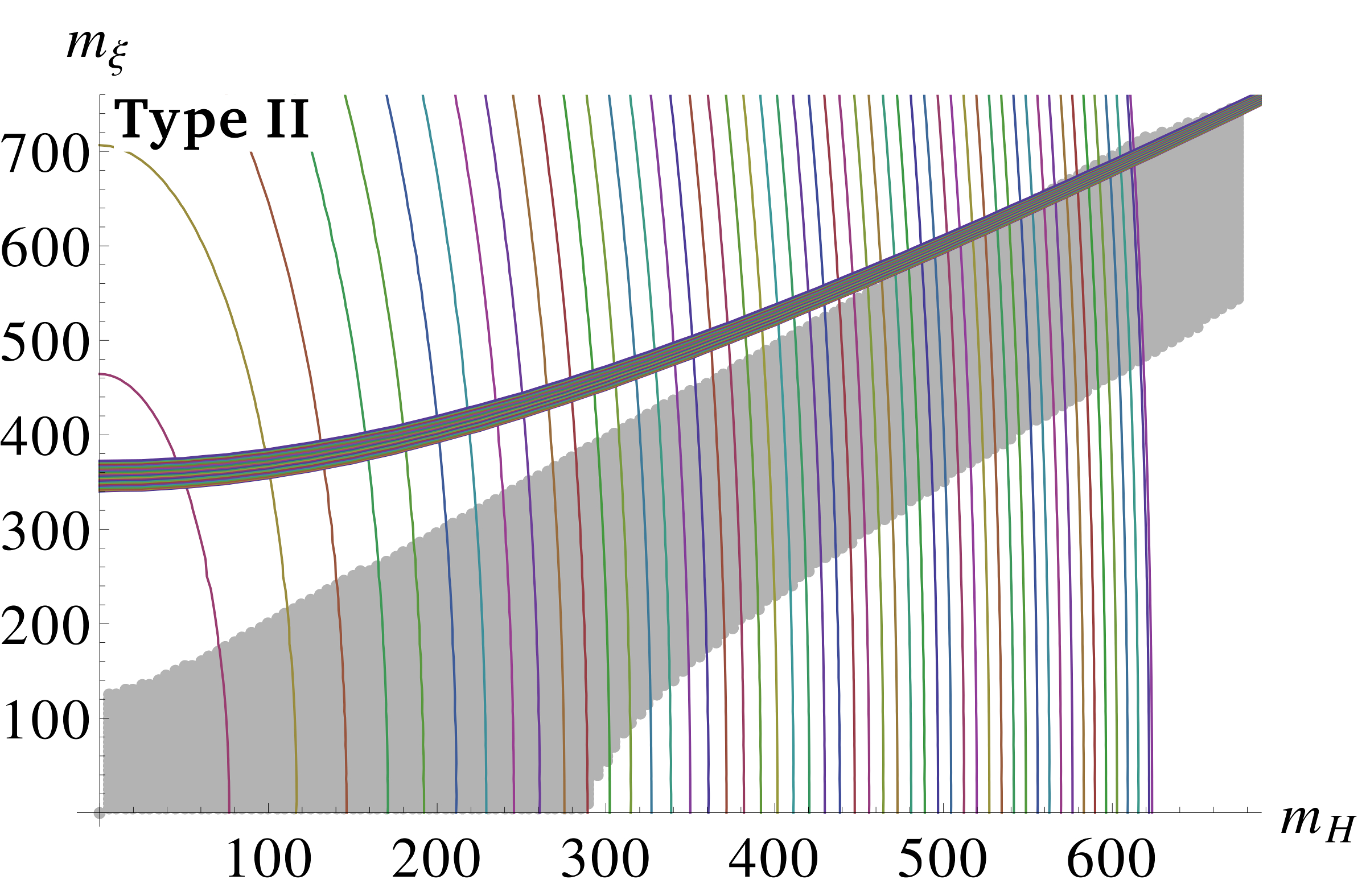}
 \label{type2}}
\subfigure[]{
 \includegraphics[height=0.3\columnwidth, width = 0.45\columnwidth]{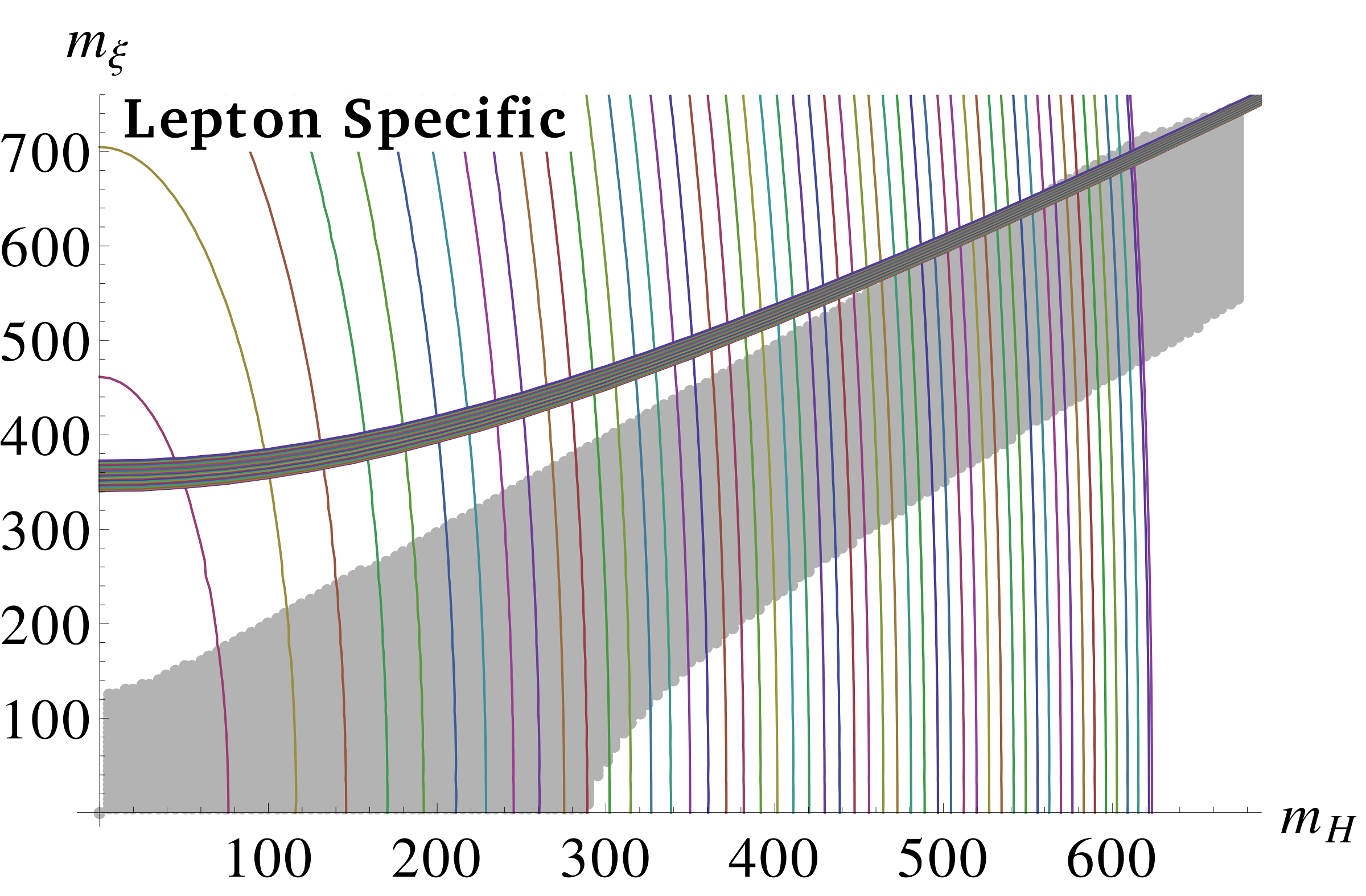}
 \label{LS}}
\subfigure[]{
 \includegraphics[height=0.3\columnwidth, width = 0.45\columnwidth]{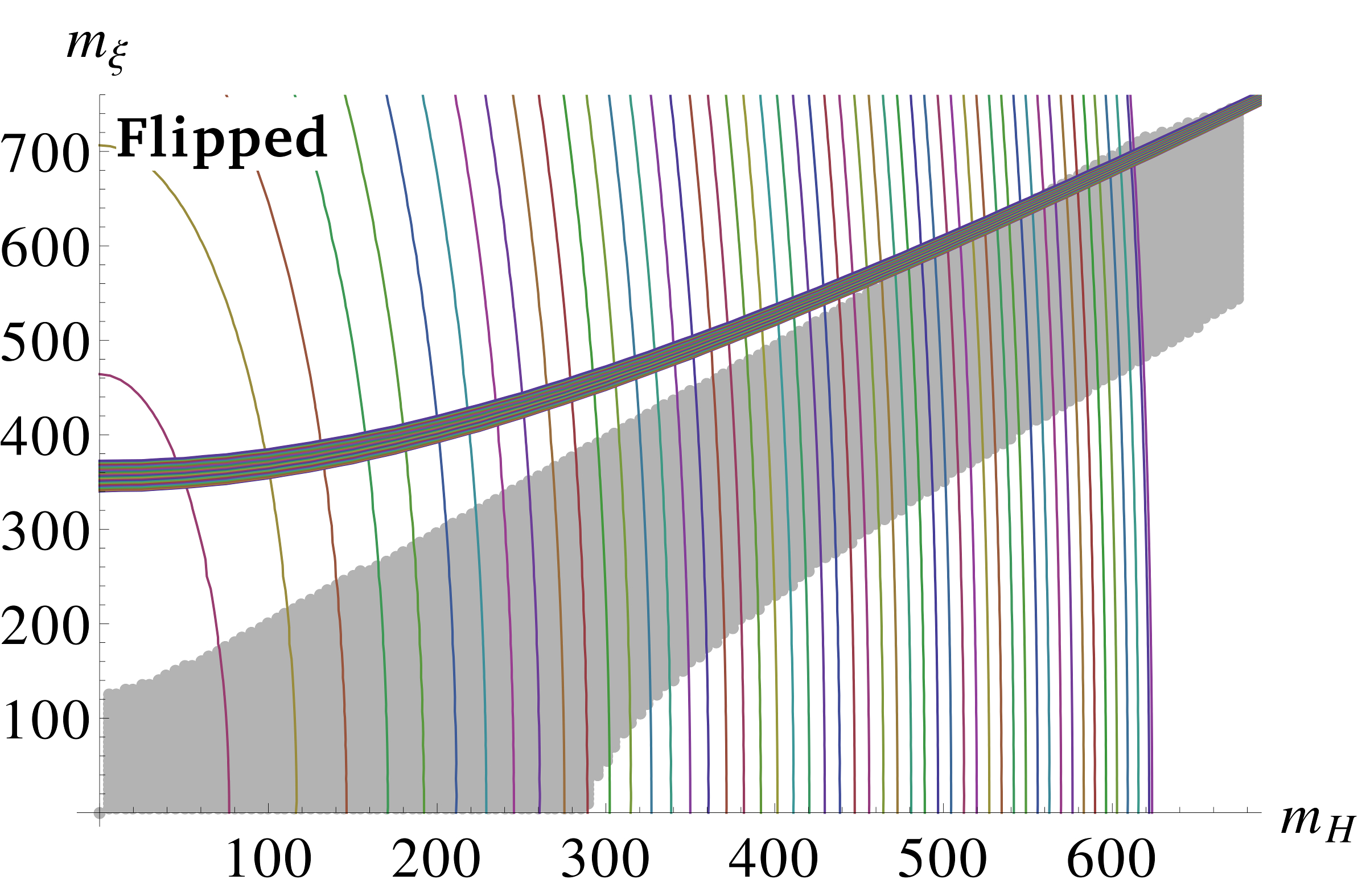}
 \label{Flipped}}
\caption{Allowed mass range (in GeV) for the charged Higgs and 
the heavy CP even Higgs in (a) type I (b) type II (c) lepton specific and
(d) flipped 2HDM 
for $\vert\lambda_{5}
 \vert \leq 4\pi$ and $\tan \beta = 5$}  
\label{fig.result}
\end{figure}

Our results consist of the pairs $(m_H, m_\xi)\,$ for each type of
2HDM, satisfying the two Veltman conditions, and consistent with
the constraints from stability, tree-level unitarity and the $T$
parameter.  For $\tan\beta = 5\,,$ we have plotted the $m_H-m_\xi$
curves corresponding to VC1 and VC2
for several values of $\lambda_5\,.$ These have been superimposed
on the bound determined by 
(\ref{mass.ineq1}), (\ref{mass.ineq2}), and
(\ref{Tvalue}).  The resulting plot is shown in
Fig.~\ref{fig.result}.  VC1 produces ellipses, and VC2 gives a
narrow band of hyperbolae.  Their crossings which fall inside the
band representing the bound from the inequalities
are the allowed masses. From the plot we can estimate the
individual bounds: for all four models, we find approximately 550
GeV $\lesssim m_\xi \lesssim$ 700 GeV, and about 450 GeV $\lesssim
m_H \lesssim$ 620 GeV, with a higher $m_H$ implying a higher
$m_\xi\,.$ As mentioned earlier, $m_A$ is close to $m_H$ as a
result of~(\ref{mass.ineq1}). We also note that direct searches have 
put a rough lower bound of 
$m_{\xi}>100$ GeV~\cite{Beringer:1900zz}.
%
\section{Discussion}
Some comments are in order for the values of some parameters that we have 
used in this analysis. We chose $\beta - \alpha = \frac\pi2$ so that the 2HDMs are
in the alignment limit, in which the lighter CP-even scalar $h$ has the couplings 
of the Higgs particle of the Standard Model. We note that in the decoupling 
limit~\cite{Gunion:2002zf} defined by $m_A^2 \gg |\lambda_i| v^2$ subject 
to a condition of perturbativity $|\lambda_i| \lesssim 4\pi$, we also find  
$\sin(\beta-\alpha)\approx 1\,$. (The relation between these $\lambda_i$ and
ours may be found in~\cite{Gunion:2002zf}.) Although we find from our computations 
in this paper that $m_A$ must be large, we do not require it a priori, so our 
results are valid for the SM-like alignment limit of the 2HDMs, without 
going to the decoupling limit. It is worth pointing out that the issue of
distinguishing between the decoupling limit and the SM-like scenario was first
explored in~\cite{Ginzburg:2004vp}. 

Perturbativity requires that the quartic couplings of the physical Higgs fields
are small. Our choice of $|\lambda_i|\leq 4\pi$ keeps the models inside the
perturbative regime, and this requirement also keeps $m_A \lesssim$ 617 GeV.
Allowing for larger values of $\lambda_i$ would also allow higher values of
$m_A$ as well as of $m_H$ and $m_\xi$\,. In that sense, what we have found
in this paper are the lower bounds on the masses of $H, A, $ and $\xi^\pm\,,$
in the SM-like limit of 2HDMs.

The most important parameter in the 2HDMs is $\tan\beta\,.$ There
is no consensus on the value of $\tan\beta\,,$ except that it
should be larger than unity, based on constraints coming from $Z\to
b\bar b$ and $B_q \bar B_q$ mixing~\cite{Arhrib:2009hc}. Several
arguments have been proffered for a large $\tan\beta\,$ in 2HDMs 
of different types, using muon $g-2$ in lepton specific 2HDM~\cite{Cao:2009as}, 
or using $b\to s\gamma$ in type I and flipped models~\cite{Park:2006gk},
which also suppresses the $t\rightarrow bH^{+}$ branching ratio
to a rough agreement with 95$\%$ CL limits from the
light charged Higgs searches at the LHC~\cite{Aad:2012tj,
	Chatrchyan:2012vca}. A large value of $\tan\beta$ also
 makes the heavy
Higgs particle difficult to detect~\cite{Randall:2007as}.  We have
used a conservative $\tan\beta= 5\,$ to estimate the scalar masses
$m_H$ and $m_{\xi^\pm}$ --- note that $m_A$ is not very far from
$m_H$ because of the degeneracy relation~(\ref{mass.ineq1}).  A
larger $\tan\beta$ makes the $m_H-m_A$ degeneracy more 
pronounced, so the inequality band becomes narrower. This narrows the
ranges of $m_H$ and $m_\xi\,,$ also pushing the region of overlap
upwards, making the heavy and charged Higgses more difficult to
detect. Recent analyses of LHC data at $\sqrt s = 8$~TeV, as a search 
for the pseudoscalar Higgs particle, also appear 
to favor a value of 5 or larger for $\tan\beta\,$ near the alignment 
limit~\cite{Aad:2015wra, Khachatryan:2015lba} .

\begin{acknowledgements}
We thank B.~Grzadkowski, B.~Swiezewska and X.-F.~Han for useful comments,
and the anonymous referee for informing us about Ref.~\cite{Ginzburg:2004vp}. 
AB wishes to thank D.~Das for useful discussions. 
\end{acknowledgements}

      

\begin{thebibliography}{99}

\bibitem{Aad:2012tfa} 
  G.~Aad {\it et al.}  [ATLAS Collaboration],
  \textit{``Observation of a new particle in the search for the Standard 
Model Higgs boson with the ATLAS detector at the LHC,''}
  Phys.\ Lett.\ B {\bf 716}, 1 (2012).

\bibitem{Chatrchyan:2012ufa} 
  S.~Chatrchyan {\it et al.}  [CMS Collaboration],
 \textit{ ``Observation of a new boson at a mass of 125 GeV with 
the CMS experiment at the LHC,''}
  Phys.\ Lett.\ B {\bf 716}, 30 (2012).

\bibitem{Branco:2011iw} 
  G.~C.~Branco, P.~M.~Ferreira, L.~Lavoura, M.~N.~Rebelo, M.~Sher 
  and J.~P.~Silva, 
 \textit{ ``Theory and phenomenology of two-Higgs-doublet models,''}
  Phys.\ Rept.\  {\bf 516}, 1 (2012)

\bibitem{Peccei:1977hh} 
  R.~D.~Peccei and H.~R.~Quinn,
 \textit{ ``CP Conservation in the Presence of Instantons,''
}  Phys.\ Rev.\ Lett.\  {\bf 38}, 1440 (1977).

\bibitem{Peccei:1977ur} 
  R.~D.~Peccei and H.~R.~Quinn,
 \emph{ ``Constraints Imposed by CP Conservation in the Presence of
  Instantons,'' }
  Phys.\ Rev.\ D {\bf 16}, 1791 (1977).

\bibitem{Turok:1990zg} 
  N.~Turok and J.~Zadrozny,
 \textit{ ``Electroweak baryogenesis in the two doublet model,''
 } Nucl.\ Phys.\ B {\bf 358}, 471 (1991).

\bibitem{Ma:2006km} 
  E.~Ma,
  \textit{``Verifiable radiative seesaw mechanism of neutrino mass and
  dark matter,''} 
Phys.\ Rev.\ D {\bf 73}, 077301 (2006).

\bibitem{Ma:2008uza} 
  E.~Ma,
 \textit{ ``Utility of a Special Second Scalar Doublet,''}
  Mod.\ Phys.\ Lett.\ A {\bf 23}, 647 (2008).
  
\bibitem{Barbieri:2006dq} 
  R.~Barbieri, L.~J.~Hall and V.~S.~Rychkov,
 \textit{ ``Improved naturalness with a heavy Higgs: 
 	An Alternative road to LHC physics,''}
  Phys.\ Rev.\ D {\bf 74}, 015007 (2006)
  [hep-ph/0603188].

\bibitem{Ferreira:2009jb} 
  P.~M.~Ferreira and D.~R.~T.~Jones,
  \textit{``Bounds on scalar masses in two Higgs doublet models,''}
  JHEP {\bf 0908}, 069 (2009)
  [arXiv:0903.2856 [hep-ph]].
\cite{Aad:2015wra}
\bibitem{Lee:1973iz} 
  T.~D.~Lee,
 \textit{ ``A Theory of Spontaneous T Violation,''}
  Phys.\ Rev.\ D {\bf 8}, 1226 (1973).

\bibitem{Gunion:1989we} 
  J.~F.~Gunion, H.~E.~Haber, G.~L.~Kane and S.~Dawson,
 \textit{``The Higgs Hunter's Guide,"}
  Front.\ Phys.\  {\bf 80}, 1 (2000).

\bibitem{Gunion:1992hs} 
  J.~F.~Gunion, H.~E.~Haber, G.~L.~Kane and S.~Dawson,
  \textit{``Errata for the Higgs hunter's guide,"}
  arXiv:hep-ph/9302272.

\bibitem{Branco:1996bq} 
  G.~C.~Branco, W.~Grimus and L.~Lavoura,
  \textit{``Relating the scalar flavor changing neutral couplings to the
  CKM matrix,''}  
Phys.\ Lett.\ B {\bf 380}, 119 (1996)

\bibitem{Gunion:2002zf} 
  J.~F.~Gunion and H.~E.~Haber,
  \textit{``The CP conserving two Higgs doublet model: The Approach to the
  decoupling limit,'' }
  Phys.\ Rev.\ D {\bf 67}, 075019 (2003)

\bibitem{Ferreira:2014naa} 
  P.~M.~Ferreira, J.~F.~Gunion, H.~E.~Haber and R.~Santos,
  Phys.\ Rev.\ D {\bf 89}, no. 11, 115003 (2014)
  [arXiv:1403.4736 [hep-ph]].
 

\bibitem{Veltman:1980mj} 
  M.~J.~G.~Veltman,
  \textit{``The Infrared - Ultraviolet Connection,''}
  Acta Phys.\ Polon.\ B {\bf 12}, 437 (1981).

 \bibitem{Newton:1993xc} 
  C.~Newton and T.~T.~Wu,
  \textit{``Mass relations in the two Higgs doublet model from the absence
  of quadratic divergences,''} 
  Z.\ Phys.\ C {\bf 62}, 253 (1994).

\bibitem{Glashow:1976nt} 
  S.~L.~Glashow and S.~Weinberg,
  \textit{``Natural Conservation Laws for Neutral Currents,''}
  Phys.\ Rev.\ D {\bf 15}, 1958 (1977).

\bibitem{Paschos:1976ay} 
  E.~A.~Paschos,
  \textit{``Diagonal Neutral Currents,''}
  Phys.\ Rev.\ D {\bf 15}, 1966 (1977).

\bibitem{Pich:2009sp} 
  A.~Pich and P.~Tuzon,
  \textit{``Yukawa Alignment in the Two-Higgs-Doublet Model,''}
  Phys.\ Rev.\ D {\bf 80}, 091702 (2009).
  
\bibitem{Ferreira:2010xe} 
P.~M.~Ferreira, L.~Lavoura and J.~P.~Silva,
 \textit{``Renormalization-group constraints on Yukawa 
alignment in multi-Higgs-doublet models,''}
Phys.\ Lett.\ B {\bf 688}, 341 (2010).


  \bibitem{Grzadkowski:2009iz} 
  B.~Grzadkowski and P.~Osland,
 \textit{ ``Tempered Two-Higgs-Doublet Model,''}
  Phys.\ Rev.\ D {\bf 82}, 125026 (2010).
  
   \bibitem{Jora:2013opa} 
   R.~Jora, S.~Nasri and J.~Schechter,
   \textit{``Naturalness in a simple two Higgs doublet model,''}
   Int.\ J.\ Mod.\ Phys.\ A {\bf 28}, 1350036 (2013).
   
\bibitem{Akeroyd:2000wc} 
A.~G.~Akeroyd, A.~Arhrib and E.~M.~Naimi,
\textit{``Note on tree level unitarity in the general two Higgs doublet
model,''} 
Phys.\ Lett.\ B {\bf 490}, 119 (2000)

   
\bibitem{Beringer:1900zz} 
  J.~Beringer {\it et al.}  [Particle Data Group Collaboration],
  {\it Review of Particle Physics (RPP),}
  Phys.\ Rev.\ D {\bf 86}, 010001 (2012) and 
  2013 partial update for the 2014 edition. 

\bibitem{Cao:2009as} 
  J.~Cao, P.~Wan, L.~Wu and J.~M.~Yang,
  \textit{``Lepton-Specific Two-Higgs Doublet Model: Experimental 
Constraints and Implication on Higgs Phenomenology,''}
  Phys.\ Rev.\ D {\bf 80}, 071701 (2009).

\bibitem{Arhrib:2009hc} 
  A.~Arhrib, R.~Benbrik, C.~H.~Chen, R.~Guedes and R.~Santos,
  \textit{``Double Neutral Higgs production in the Two-Higgs doublet model
  at the LHC,'' }
  JHEP {\bf 0908}, 035 (2009).

\bibitem{Sher:1988mj} 
  M.~Sher,
 \textit{ ``Electroweak Higgs Potentials and Vacuum Stability,''}
  Phys.\ Rept.\  {\bf 179}, 273 (1989).

\bibitem{Maalampi:1991fb} 
  J.~Maalampi, J.~Sirkka and I.~Vilja,
  \textit{``Tree level unitarity and triviality bounds 
  	for two Higgs models,''}
  Phys.\ Lett.\ B {\bf 265}, 371 (1991).

\bibitem{Kanemura:1993hm} 
  S.~Kanemura, T.~Kubota and E.~Takasugi,
  \textit{``Lee-Quigg-Thacker bounds for Higgs boson masses in a two
  doublet model,''} 
  Phys.\ Lett.\ B {\bf 313}, 155 (1993).

\bibitem{Horejsi:2005da} 
  J.~Horejsi and M.~Kladiva,
  \textit{``Tree-unitarity bounds for THDM Higgs masses revisited,''}
  Eur.\ Phys.\ J.\ C {\bf 46}, 81 (2006).

\bibitem{Bhattacharyya:2013rya} 
  G.~Bhattacharyya, D.~Das, P.~B.~Pal and M.~N.~Rebelo,
  \textit{``Scalar sector properties of two-Higgs-doublet models with a
  global U(1) symmetry,'' }
  JHEP {\bf 1310}, 081 (2013).


\bibitem{He:2001tp} 
  H.~J.~He, N.~Polonsky and S.~f.~Su,
  \textit{``Extra families, Higgs spectrum and oblique corrections,''}
  Phys.\ Rev.\ D {\bf 64}, 053004 (2001).

\bibitem{Grimus:2007if} 
  W.~Grimus, L.~Lavoura, O.~M.~Ogreid and P.~Osland,
  \textit{``A Precision constraint on multi-Higgs-doublet models,''}
  J.\ Phys.\ G {\bf 35}, 075001 (2008).

\bibitem{Baak:2013ppa} 
  M.~Baak and R.~Kogler,
  {\it ``The global electroweak Standard Model fit after the Higgs
  discovery''}\,, 
  arXiv:1306.0571 [hep-ph].
  
\bibitem{Ginzburg:2004vp} 
I.~F.~Ginzburg and M.~Krawczyk,
\textit{``Symmetries of two Higgs doublet model and CP violation,''}
Phys.\ Rev.\ D {\bf 72}, 115013 (2005)

\bibitem{Park:2006gk} 
  J.~h.~Park,
  \textit{``Lepton non-universality at LEP and charged Higgs,''}
  JHEP {\bf 0610}, 077 (2006).

\bibitem{Aad:2012tj} 
  G.~Aad {\it et al.}  [ATLAS Collaboration],
  \textit{``Search for charged Higgs bosons decaying via $H^{+} \to \tau \nu$ 
in top quark pair events using $pp$ collision data at $\sqrt{s}=7$ TeV 
with the ATLAS detector,''}
  JHEP {\bf 1206}, 039 (2012).

\bibitem{Chatrchyan:2012vca} 
  S.~Chatrchyan {\it et al.}  [CMS Collaboration],
 \emph{ ``Search for a light charged Higgs boson in top quark decays in $pp$ collisions at $\sqrt{s}=7$ TeV,''}
  JHEP {\bf 1207}, 143 (2012).
  
\bibitem{Randall:2007as} 
  L.~Randall,
 \emph{ ``Two Higgs Models for Large Tan Beta and Heavy Second Higgs,''}
  JHEP {\bf 0802}, 084 (2008).
  
\bibitem{Aad:2015wra} 
G.~Aad {\it et al.}  [ATLAS Collaboration],
\textit{``Search for a CP-odd Higgs boson decaying to Zh in pp collisions at $\sqrt{s} = 8$ TeV with the ATLAS detector,''}
Phys.\ Lett.\ B {\bf 744}, 163 (2015).

\bibitem{Khachatryan:2015lba} 
V.~Khachatryan {\it et al.}  [CMS Collaboration],
\textit{``Search for a pseudoscalar boson decaying into a Z boson and the 125 GeV Higgs boson in llbb final states,''}
arXiv:1504.04710 [hep-ex].


\end{thebibliography}
\end{document}